\documentclass[%
 reprint,superscriptaddress,nofootinbib
 amsmath,amssymb,
 aps,
]{revtex4-1}
\usepackage[sort&compress]{natbib}
\usepackage{graphicx}
\usepackage{dcolumn}
\usepackage{bm}
\usepackage{float}
\usepackage{changes}
\usepackage{color}
\usepackage{varioref}
\usepackage{hyperref}
\usepackage{cleveref}
\hypersetup{
    colorlinks,
    citecolor=black,
    filecolor=black,
    linkcolor=black,
    urlcolor=black
}
\def\vecsign{\mathchar"017E}
\def\dvecsign{\smash{\stackon[-1.95pt]{\vecsign}{\rotatebox{180}{$\vecsign$}}}}
\def\dvec#1{\def\useanchorwidth{T}\stackon[-4.2pt]{#1}{\,\dvecsign}}
\usepackage{stackengine}
\stackMath
\usepackage{textcomp}

\begin{document}

\preprint{APS/123-QED}

\title{Optical Magnetism in Planar Metamaterial Heterostructures}

\author{Georgia T. Papadakis}\email{gpapadak@caltech.edu}
\affiliation{Thomas J. Watson Laboratories of Applied Physics, California Institute of Technology, California 91125, USA}
\author{Dagny Fleischman}
\affiliation{Thomas J. Watson Laboratories of Applied Physics, California Institute of Technology, California 91125, USA}
\affiliation{Kavli Nanoscience Institute, California Institute of Technology, California 91125, USA}
\author{Artur Davoyan}
\affiliation{Thomas J. Watson Laboratories of Applied Physics, California Institute of Technology, California 91125, USA}
\affiliation{Kavli Nanoscience Institute, California Institute of Technology, California 91125, USA}
\affiliation{Resnick Sustainability Institute, California Institute of Technology, California 91125, USA}
\author{Pochi Yeh}
\affiliation{Department of Electrical and Computer Engineering, University of Santa Barbara, California 93106, USA}
\author{Harry A. Atwater}
\affiliation{Thomas J. Watson Laboratories of Applied Physics, California Institute of Technology, California 91125, USA}

\date{\today}

\begin{abstract}
  \textcolor{black}{Harnessing artificial optical magnetism has required rather complex two- and 
three-dimensional structures, examples include split-ring and fishnet metamaterials 
and nanoparticles with non-trivial magnetic properties. By contrast, dielectric 
properties can be tailored even in planar and pattern-free, one-dimensional (1D) 
arrangements, for example metal/dielectric multilayer metamaterials. 
These systems are extensively investigated due to their hyperbolic and plasmonic response, which, however, has been 
previously considered to be limited to transverse magnetic (TM) polarization, based on the general consensus that they do
 not possess interesting magnetic properties.
In this work, we tackle these two seemingly unrelated issues 
  simultaneously, by proposing conceptually and demonstrating experimentally a mechanism for artificial magnetism 
  in planar, 1D metamaterials.
  We show experimentally that the magnetic response of metal/high-index dielectric hyperbolic metamaterials 
  can be anisotropic, leading to frequency regimes 
  of magnetic hyperbolic dispersion. We investigate the implications of our results for transverse electric (TE) polarization and 
  show that such systems can support TE interface-bound states, analogous
   to their TM counterparts, surface plasmon polaritons. Our results simplify the structural 
   complexity for tailoring artificial magnetism in lithography-free 
  systems and generalize the concept of plasmonic and hyperbolic properties to encompass 
  both TE and TM polarizations at optical frequencies.}
\begin{description}
\item[PACS numbers]
78.67.Pt, 73.20.Mf, 75.30.Gw, 78.20.Ls\\
\end{description}
\end{abstract}

\pacs{Valid PACS appear here}
\maketitle


\section{\label{sec:level1}INTRODUCTION}

\par{In the optical spectral range, the magnetic response of most materials, given by the magnetic permeability $\mu$, 
is generally weak. This is famously expressed in the textbook by Landau and Lifshitz \cite{Landau_Lifshitz}: 
``\textit{there is no meaning
 in using the magnetic susceptibility from the optical frequencies onward, and in discussing such phenomena, 
 we must put $\mu=1$}''. The magnetic properties of natural materials arise from microscopic orbital currents and 
 intrinsic spins and typically vanish at frequencies above the GHz range. This has motivated a search for 
 structures and systems that may exhibit artificial optical magnetism by utilizing principles of metamaterial
  design. In this regime, engineered displacement and conduction currents, induced when metamaterials 
  are illuminated with electromagnetic fields, act as sources of artificial magnetism \cite{Monticone_Magnetism}.}

\par{Maxwell\textquotesingle s equations exhibit a duality with respect to dielectric permittivity $\epsilon$ and magnetic permeability $\mu$ 
for the two linear polarizations of light; transverse magnetic (TM) and transverse electric (TE), respectively. 
Despite this symmetry, at frequencies beyond THz, far more discussion in the literature has been devoted 
to tailored dielectric properties of metamaterials than to artificial magnetic properties.  This imbalance is 
understandable because, until now, \textcolor{black}{the realization of} magnetic metamaterials has required rather complex resonant 
geometries \cite{Monticone_Magnetism, Shalaev_Negativeindex, Brener_DielectricMagnetic}, such as arrays of paired thin 
metallic strips \cite{Shalaev_Metallicstrips, Shvetz_Metallicstrips}, split 
ring resonators \cite{Kafesaki_Splitring, Giessen_Splitring, Aydin_SRR} or fishnet structures 
\cite{Kruk_Fishnet}, which are challenging to realize experimentally at optical frequencies.}

\par{In contrast, engineered dielectric 
properties of metamaterials are achievable even in simple planar multilayer configurations. In fact, heterostructures of 
alternating metallic and dielectric layers, termed “hyperbolic metamaterials” (HMMs), have been 
explored intensively the last decade \cite{Kivshar_HMMReview, Drachev_HMM, Krishnamoorthy_TopTransitions}. They 
are often described with an effective permittivity tensor, $\dvec{\epsilon}_{\mathrm{eff}}=diag\{\epsilon_{\mathrm{o}},\epsilon_{\mathrm{o}},\epsilon_{\mathrm{e}}\}$, 
where the subscript-o (e) indicates the ordinary (extraordinary) direction. Their in-plane response is metallic 
($\epsilon_{\mathrm{o}}<0$) while their out-of-plane response is dielectric ($\epsilon_{\mathrm{e}}>0$). HMMs support 
interesting electromagnetic phenomena, including negative refraction \cite{Kivshar_HMMReview, High_Hyperbolicmetasurface} 
without the need of a negative refractive index, diverging density of optical states for Purcell-factor 
enhancement \cite{Krishnamoorthy_TopTransitions}, and hyper-lensing \cite{Lu_Hyperlens}. Furthermore, 
the negative dielectric permittivity $\epsilon_o$ leads to surface-propagating plasmonic modes 
\cite{Elser_AnisotropicPlasmon, Narimanov_Hypercrystal}, similar to surface 
plasmon polaritons (SPPs) supported on noble 
metals' surfaces \cite{Economou_SPP} or metal/dielectric arrangements and 
waveguides \cite{Maier_Plasmonics, Lezec_NegativeRefraction, Dionne_NegativeIndex, Fang_TammMetalDiel}.}

\par{The plasmonic and hyperbolic properties of planar, multilayer heterostructures 
have featured prominently in photonics, however, their relevance has been limited to 
TM polarization, based on their effective dielectric response. \textcolor{black}{TE polarization-related 
phenomena have remained unexplored, as the effective magnetic permeability of such systems has been 
widely considered to be unity \cite{Kivshar_HMMReview, Drachev_HMM, Krishnamoorthy_TopTransitions, 
Jacob_DOS}. By contrast, fishnet metamaterials are known to exhibit a magnetic 
response and, recently, a magnetic hyperbolic metamaterial was demonstrated 
\cite{Kruk_Fishnet}. However, the fishnet structure is by definition biaxial, thus, 
TE and TM polarizations cannot be independently manipulated.}}

\par
\begin{figure}[H]
 \centering
 \includegraphics[width=0.95\linewidth]{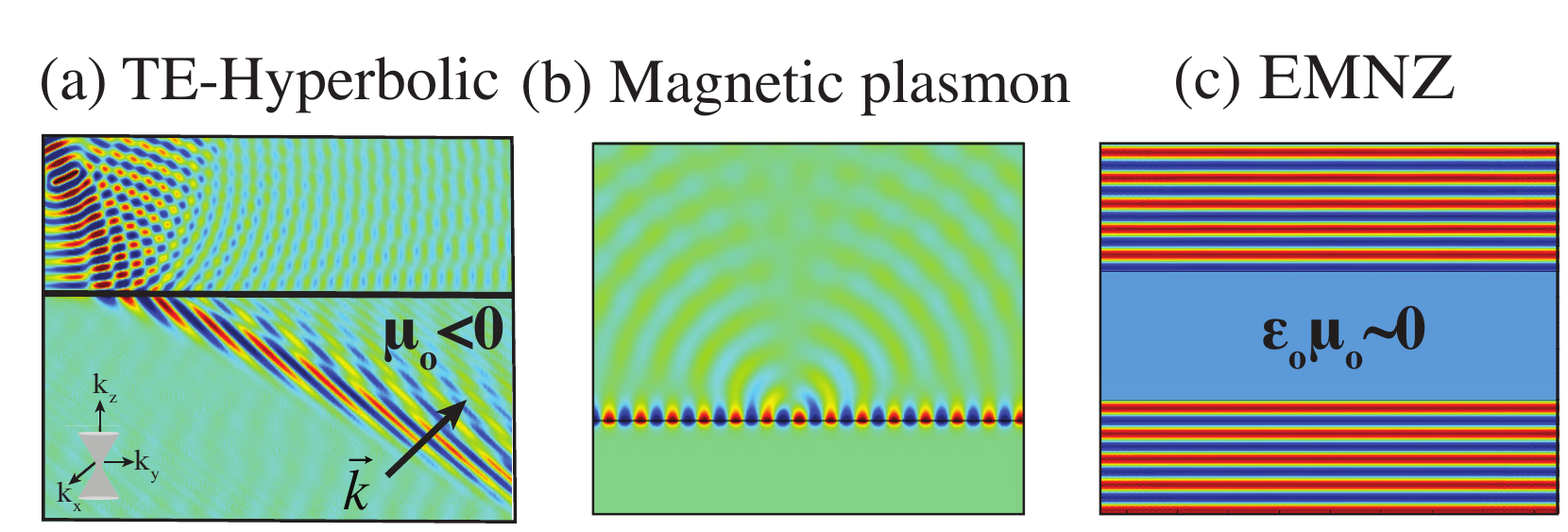}
  \caption{(a) TE hyperbolic refraction in type II HMMs ($\mu_{\mathrm{o}}<0$, $\mu_{\mathrm{e}}>0$)-inset: 3D isofrequency diagram 
  (b) TE magnetic plasmon: TE polarized surface state at the interface between air and magnetic material
   ($\mu<0$), analogous to TM polarized surface plasmon polaritons ($\epsilon<0$). (c) $\epsilon$ and $\mu$ near zero (EMNZ) 
   regime: phase diagram demonstrating vanishing phase advance at EMNZ wavelengths.}
 \label{fig:Figure1}
\end{figure}

{\textcolor{black}{Here, we focus on the magnetic properties of unpatterned, one-dimensional (1D) multilayer uniaxial 
systems, where TE and TM polarizations are uncoupled. Inducing an artificial 
magnetic response in these systems is of interest for generalizing their 
hyperbolic and plasmonic properties to encompass both TE and TM polarizations. 
For example, a metal/dielectric planar system with opposite magnetic 
permeabilities along different coordinate directions ($\mu_{\mathrm{o}}\mu_{\mathrm{e}}<0$) is double hyperbolic with 
unbound wavevectors for TE polarization (Fig. \ref{fig:Figure1}a and 
inset).} Furthermore, no TE counterpart of the 
surface plasmon polariton, i.e. a magnetic surface plasmon (Fig. \ref{fig:Figure1}b), has been reported 
at optical frequencies, due to lack of negative magnetic response. Artificial 
epsilon-and-mu-near-zero (EMNZ) metamaterials at optical frequencies are interesting building blocks
 for electrostatic-like systems, due to near-zero phase advance in the 
 material \cite{Mahmoud_EMNZ} (Fig. \ref{fig:Figure1}c). While it is straightforward to tailor the 
 permittivity to cross zero in planar metamaterials \cite{Maas_ENZ}, a simultaneously EMNZ metamaterial 
 at optical frequencies has not yet been demonstrated.}
 
\par{\textcolor{black}{In this paper, we perform a comprehensive study of 
artificial magnetism in planar, 1D multilayer metamaterials. 
The practical importance of our results lies in the drastic simplification of the structural complexity
 of previous generation magnetic metamaterials; the realization of split-ring 
 resonators \cite{Kafesaki_Splitring, Aydin_SRR, Giessen_Splitring}, fishnet 
structures \cite{Kruk_Fishnet}, and nanoparticles \cite{Kuznetsov_MagneticLight, Alaeian_Magnetic} at optical 
frequencies requires multi-step lithographic processes and synthesis. By contrast, \textit{pattern-free} 
multilayer metamaterials are readily realizable with lithography-free thin-film 
deposition. We start by introducing the physical concept for inducing an artificial magnetic response 
in 1D systems (Sec.\ref{Concept}). In Sec.\ref{Relaxing the mu condition}, we briefly discuss
 a simple approach based on which this magnetic response can be taken into account in the design of 
 multilayer metamaterials by relaxing previously made assumptions in widely used effective medium 
theories \cite{Agranovich_EMA, Ni_Bloch, Krishnamoorthy_TopTransitions, Kivshar_HMMReview, 
Drachev_HMM}. In Sec. \ref{Experimental Results}, we experimentally confirm our findings and 
demonstrate magnetic resonances at optical frequencies in multilayer HMMs.  
\textcolor{black}{Motivated by the non-trivial effective magnetic response that we 
observe experimentally, in Sec. \ref{Implications} we theoretically investigate 
its implications using a simple transfer-matrix approach. Contrary to the majority of work in planar plasmonics 
 and HMMs, we investigate TE polarization phenomena. We find that concepts previously discussed for TM 
 polarization, based on engineering the dielectric permittivity, are generalizable for both linear 
  polarizations. The proposed 
 effective description of 1D systems in terms of effective dielectric \textit{and} 
 magentic parameters provides a simple and intuitive understanding of the underlying physics.}}
  
\vspace{-1.0em}
\section{\label{Concept}PHYSICAL MECHANISM: INDUCED MAGNETIC DIPOLES IN 1D SYSTEMS}

\par{\textcolor{black}{Magnetic fields at radio frequencies are usually manipulated with induction 
coils that generate and induce magnetic flux. They operate based on circulating conduction 
 currents in coil loops that can be approximated as magnetic dipoles. This concept is widespread in
  metamaterials design \cite{Smith_Coil, Capolino_Book}, where the conduction current is often replaced
   by displacement current in artificially magnetic structures at higher 
   frequencies. Similar to the RF regime, by properly shaping metamaterial elements to produce 
 a circulating current flow, magnetic dipoles are induced.
 Dielectric nanoparticles \cite{Engheta_Nanocapacitors, Alaeian_Magnetic, Jacob_DielectricMMs, Kuznetsov_MagneticLight, Evlyukhin_MagneticNanosphere, Schilling_Sinanophotonics}
 and nanorods \cite{Mirmoosa_MagneticHMM, OBrien_MetallicColumns} have been the building blocks 
 for three (3D)- and two (2D)-dimensional magnetic metamaterial structures,  
 respectively (Fig. \ref{fig:Figure2}a, b). In both cases, the circular geometry allows for loop-like current 
 flow, generating a magnetic moment.} \textcolor{black}{We note that 
 the magnetic response of these arrangements is sometimes incorporated into an equivalent, alternative, spatially dispersive
  permittivity. Although this is, in principle, always possible \cite{Alu_SpatialDispersionEquivalent, Landau_Lifshitz, 
  Agranovich_SpatialDispersion}, we stress that, similar to naturally occuring substances, described with a
   permittivity $\epsilon$ and a permeability $\mu$, a metamaterial description based on ($\epsilon$, $\mu$) allows for
   physical intuition and reduces complexity, especially when it is straightforward to relate the dielectric (magnetic) response with 
   physical macroscopic electric (magnetic) moments. This can be particularly useful for uniaxial 
   planar and unpatterned 1D multilayers, as, in this case, TE and TM linear
    polarizations are decoupled and directly associated with $\mu$ and $\epsilon$, respectively.}}

\begin{figure}[ht!]
 \centering
 \includegraphics[width=0.8\linewidth]{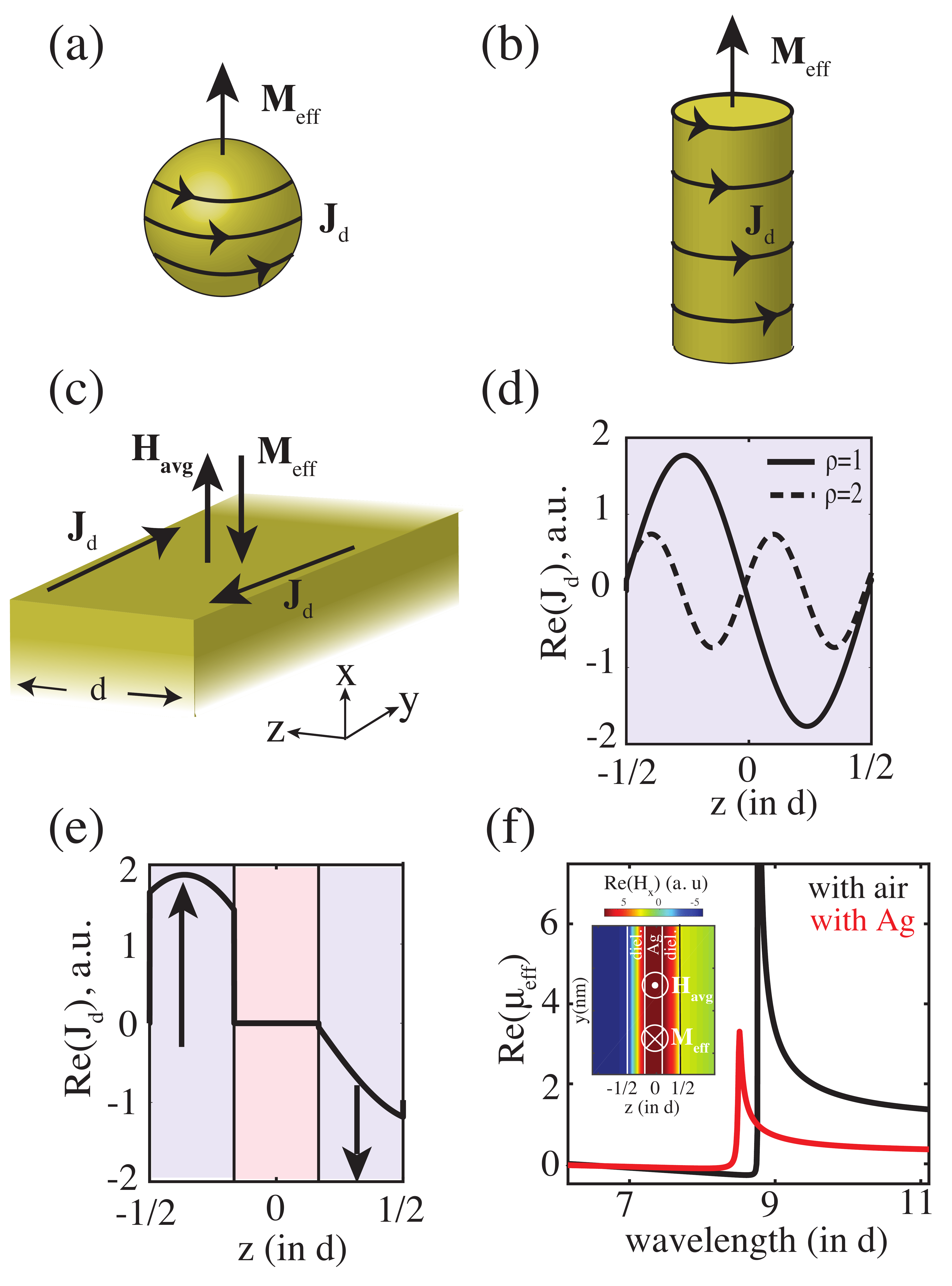}
  \caption{Induced magnetization in (a) dielectric nanoparticles (three-dimensional metamaterials) (b) 
  in dielectric nanorods (two-dimensional metamaterials) and (c) in a one-dimensional dielectric slab. 
  (d) Displacement current distribution at resonance, for $\rho=1$, $\rho=2$ for a 90nm slab of refractive index
   $n_{\mathrm{diel}}=4.5$. (e) Displacement current distribution for two dielectric layers separated by air. (f) Effective permeability 
  for two dielectric layers separated by air-black and silver-red. Inset: tangential magnetic field profile at 
  resonance: average magnetic field is opposite to $M_{\mathrm{eff}}$.}
 \label{fig:Figure2}
\end{figure}

\par{\textcolor{black}{Here, we demonstrate, a principle for strong magnetic response in 
1D layered metamaterial heterostructures. }We start by considering a single 
subwavelength dielectric slab of refractive index $n_{\mathrm{diel}}$  and thickness $d$.
When illuminated at normal incidence ($z$ direction in Fig. \ref{fig:Figure2}c), its displacement current 
$\vec{J_{\mathrm{d}}}=i\omega\epsilon_o(n_{\mathrm{diel}}^2-1)\vec{E}$ induces a macroscopic effective magnetization 
$\vec{M_{\mathrm{eff}}}=1/2\mu_o\int (\vec{r}\times\vec{J_{\mathrm{d}}})\cdot\vec{dS}$ \cite{Landau_Lifshitz, Mirmoosa_MagneticHMM, Pendry_Homogenization}. 
By averaging the magnetic field, $\langle H \rangle=\int_{-d/2}^{d/2} {H(z)}dz$, we use $\mu_{\mathrm{eff}}\simeq1+M_{\mathrm{eff}}/(\mu_o\langle H \rangle)$
to obtain an empirical closed-form expression for the magnetic permeability:
 \begin{equation}\label{eq:1}
\mu_{\mathrm{eff}}\simeq1-\frac{n_{\mathrm{diel}}^2-1}{2n_{\mathrm{diel}}^2}\left\{\ -1+\frac{n_{\mathrm{diel}} \pi d/\lambda}{\mathrm{tan}(n_{\mathrm{diel}} \pi d/\lambda)} \right\}
\end{equation}

\textcolor{black}{By setting $n_{\mathrm{diel}}=1$, we recover the unity magnetic permeability of free space. 
From Eq.(\ref{eq:1}), we see that the magnetic permeability $\mu_{\mathrm{eff}}$ will 
diverge when $\mathrm{tan}(n_{\mathrm{diel}}\pi d/\lambda)=0$. This yields a magnetic resonant behavior at free-space 
wavelengths $\lambda=n_{\mathrm{diel}}d/\rho$, with $\rho=1,2,..$. At these wavelengths, the displacement current distribution 
is anti-symmetric, as shown in Fig. \ref{fig:Figure2}d for $\rho=1,2$.} This 
anti-symmetric current flow closes a loop in $y=\pm\infty$ and induces a magnetization  
 $\vec{M_{\mathrm{eff}}}$  which is opposite to the incoming magnetic field (Fig. \ref{fig:Figure2}c), leading to a
  diamagnetic response. Eq.(\ref{eq:1}) serves 
to estimate the design parameters for enhanced magnetic response; in the long-wavelength 
limit, only the fundamental and second resonances,
 $\lambda=n_{\mathrm{diel}}d, n_{\mathrm{diel}}d/2$, play significant roles. In the visible and 
 near infrared regime, with layer thicknesses on the order of 10-100 nm,  dielectric indices higher 
 than  $n_{\mathrm{diel}}\sim2$ are required for strong magnetic effects. The same principle applies for
  grazing incidence, with the displacement current inducing a magnetic response in the 
out-of-plane ($z$) direction.}

\par{\textcolor{black}{In order to make this magnetic response significant, we consider the
 case of two parallel metallic wires in air, carrying opposite currents; their magnetic moment scales with their 
 distance, as dictated by $\vec{M}\propto\vec{r}\times\vec{J}$. In the planar geometry 
 considered here, an equivalent scheme is represented by two high-index layers 
 separated by air, as shown in Fig. \ref{fig:Figure2}e. Indeed, 
  as demonstrated with the black curve in Fig. \ref{fig:Figure2}f, the magnetic permeability 
  $\mu_{\mathrm{eff}}$ of this system strongly deviates from unity. 
  In fact, the separation layer is not required
  to be air; any high-low-high refractive index sequence will induce the same effect. 
  For example, replacing the air region with a layer of metal, with $n_{\mathrm{metal}} \ll1$ 
  at optical frequencies, does not 
  drastically change the magnetic response. This is shown for a separation layer of silver in 
  Fig. \ref{fig:Figure2}f with the red curve.} Therefore, at optical frequencies, where the conduction 
  current in metallic layers is small, metals do not contribute significantly to the magnetic response, 
  in contrast to the GHz regime, where the metallic component in resonant structures has been necessary 
  for strong magnetic effects \cite{Shvetz_Metallicstrips, Giessen_Splitring, Kafesaki_Splitring, Aydin_SRR}. 
  From the inset of Fig. \ref{fig:Figure2}f, one can see that the average magnetic field faces in the 
  direction opposite to the magnetization, expressing a negative magnetic response for the 
  dielectric/silver unit cell.}

\vspace{-1.0em}
\section{\label{Magnetic HMMs}COMBINING HYPERBOLIC DIELECTRIC AND MAGNETIC PROPERTIES}

\par{\textcolor{black}{Alternating layers of metals and dielectrics have a distinct \textit{dielectric} response, 
which is hyperbolic for TM polarization; the metallic component 
 allows for $\epsilon_{\mathrm{o}}<0$, while the dielectric layers act as barriers of conduction in the out-of-plane
  ($z$) direction, leading to $\epsilon_{\mathrm{e}}>0$.  We combine this concept with the principle for creating 
  magnetic resonances in planar systems, discussed in Sec.\ref{Concept}. We show that it is possible to induce
   an additional \textit{magnetic} response in planar dielectric/metal hyperbolic metamaterials, if the 
   dielectric layers are composed of high-index materials, capable of supporting strong displacement
    currents at optical frequencies. Previous considerations mostly pertained to 
    lower-refractive index dielectric layers, for example, LiF \cite{Tumkur_LiF} or
 Al\textsubscript{2}O\textsubscript{3} \cite{Jacob_DOS, Kim_Al2O3,  Ni_Al2O3} 
 and TiO\textsubscript{2} \cite{Krishnamoorthy_TopTransitions}. As can be inferred from 
 Fig. \ref{fig:Figure3} in what follows, for layer thicknesses below $\sim$50nm, 
 these lower-index dielectric/metal systems exhibit magnetic resonances in the ultraviolet (UV)-short wavelength 
 visible regime.}}
 
 \par{\textcolor{black}{The effective magnetic response of planar multilayer metamaterials is a 
 uniaxial tensor $\dvec{\mu}_{\mathrm{eff}}=diag\{\mu_{\mathrm{o}},\mu_{\mathrm{o}},\mu_{\mathrm{e}}\}$ and, as we demonstrate below, it 
 may also be extremely anisotropic, which has interesting implications for TE polarization
  (See Sec.\ref{Implications}). We point out, however, that the dielectric hyperbolic 
  response $\epsilon_{\mathrm{o}}\epsilon_{\mathrm{e}}<0$ in planar systems is broadband. In contrast, the magnetic 
  permeabilities along both coordinate directions $\mu_{\mathrm{o}}$ and $\mu_{\mathrm{e}}$ deviate from unity in a resonant
   manner, thus, TE polarization-based phenomena are more narrow-band in nature.}}
   
\vspace{-1.5em}
\subsection{\label{Relaxing the mu condition}Relaxing the $\mu_{\mathrm{eff}}=1$ constraint}

\par{\textcolor{black}{Prior to delving into experimental results, we briefly discuss 
our computational method, which allows relaxing the previously made $\mu_{\mathrm{eff}}=1$ 
assumption. The most extensively used approach for describing the effective 
response of hyperbolic multilayer metamaterials is the Maxwell Garnett effective medium 
approximation (EMA) \cite{Kivshar_HMMReview} (and references therein), \cite{Krishnamoorthy_TopTransitions, Drachev_HMM}, based on
 which the in-plane dielectric permittivity is given 
 by $\epsilon_{\mathrm{o,MG}}=f\epsilon_{\mathrm{m}}+(1-f)\epsilon_{\mathrm{d}}$ and the out-of-plane extraordinary 
permittivity is $\epsilon_{\mathrm{e,MG}}^{-1}=f\epsilon_{\mathrm{m}}^{-1}+(1-f)\epsilon_{\mathrm{d}}^{-1}$, where 
$f$ is the metallic filling fraction \cite{Agranovich_EMA}, while $\mu_{\mathrm{eff}}$ is \textit{a priori} set to unity. Another 
commonly used approach is the Bloch formalism, based on which, a 
periodic A-B-A-B\ldots superlattice is described with a Bloch wavenumber 
\cite{PochiYeh_Book}, which is directly translated to an effective dielectric 
permittivity \cite{Ni_Bloch}. These approaches are useful and simple to 
use, however, they are both based on the assumption of an infinite and purely periodic medium, 
without accounting for the finite thickness of realistic stacks.}}

\par{\textcolor{black}{By contrast, metamaterials other than planar ones, which are, in general, more complicated 
in structure, for example split-ring resonators \cite{Smith_SRRnegativeindex, Giessen_Splitring, Aydin_SRR, Kafesaki_Splitring}, nanoparticles 
\cite{Alaeian_Magnetic}, fishnet structures \cite{Kafesaki_FishnetVariations, Wegener_Fishnet} and many 
others, are modeled with exact S-parameter retrieval approaches \cite{Soukoulis_Retrieval, 
Chen_Retrieval}. S-parameter retrievals solve the inverse problem of determining 
the effective dielectric permittivity and magnetic permeability, $\epsilon_{\mathrm{eff}}$ and 
$\mu_{\mathrm{eff}}$ respectively, of a homogeneous slab with the same scattering 
properties, namely transmission $T$ and reflection $R$ coefficients, as the arbitrary 
inhomogeneous, composite metamaterial system of finite thickness $d$.}}

\par{\textcolor{black}{By lifting the assumption of an infinite medium, one is 
able to compute both transmission $T$ and reflection $R$ coefficients, and utilize them in S-parameter 
approaches. This allows to obtain an effective wavenumber $k_{\mathrm{eff}}$ together 
with an effective impedance $Z_{\mathrm{eff}}$ for the system under study \cite{Soukoulis_Retrieval, Chen_Retrieval}. 
These parameters are then used to decouple the effective permittivity from the permeability 
through $k_{\mathrm{eff}}=\sqrt{\epsilon_{\mathrm{eff}}\mu_{\mathrm{eff}}}\frac{\omega}{c}$ 
 and $Z_{\mathrm{eff}}=\sqrt{\frac{\mu_{\mathrm{eff}}}{\epsilon_{\mathrm{eff}}}}$. By contrast, Bloch-based 
 approaches \cite{PochiYeh_Book, Ni_Bloch} only consider a Bloch wavenumber $K_{\mathrm{Bloch}}$ (based on periodicity), 
 with no other information available for allowing decoupling $\mu_{\mathrm{eff}}$ from $\epsilon_{\mathrm{eff}}$. 
 Both the Maxwell Garnett result \cite{Agranovich_EMA} and its Bloch-based generalizations (for example
\cite{Ni_Bloch}) are based on the assumption that $\mu_{\mathrm{eff}}=1$. For a schematic comparison 
of the two approaches, see Figs.\ref{fig:Figure3}a, b.}}

\begin{figure}[]
 \centering
 \includegraphics[width=0.99\linewidth]{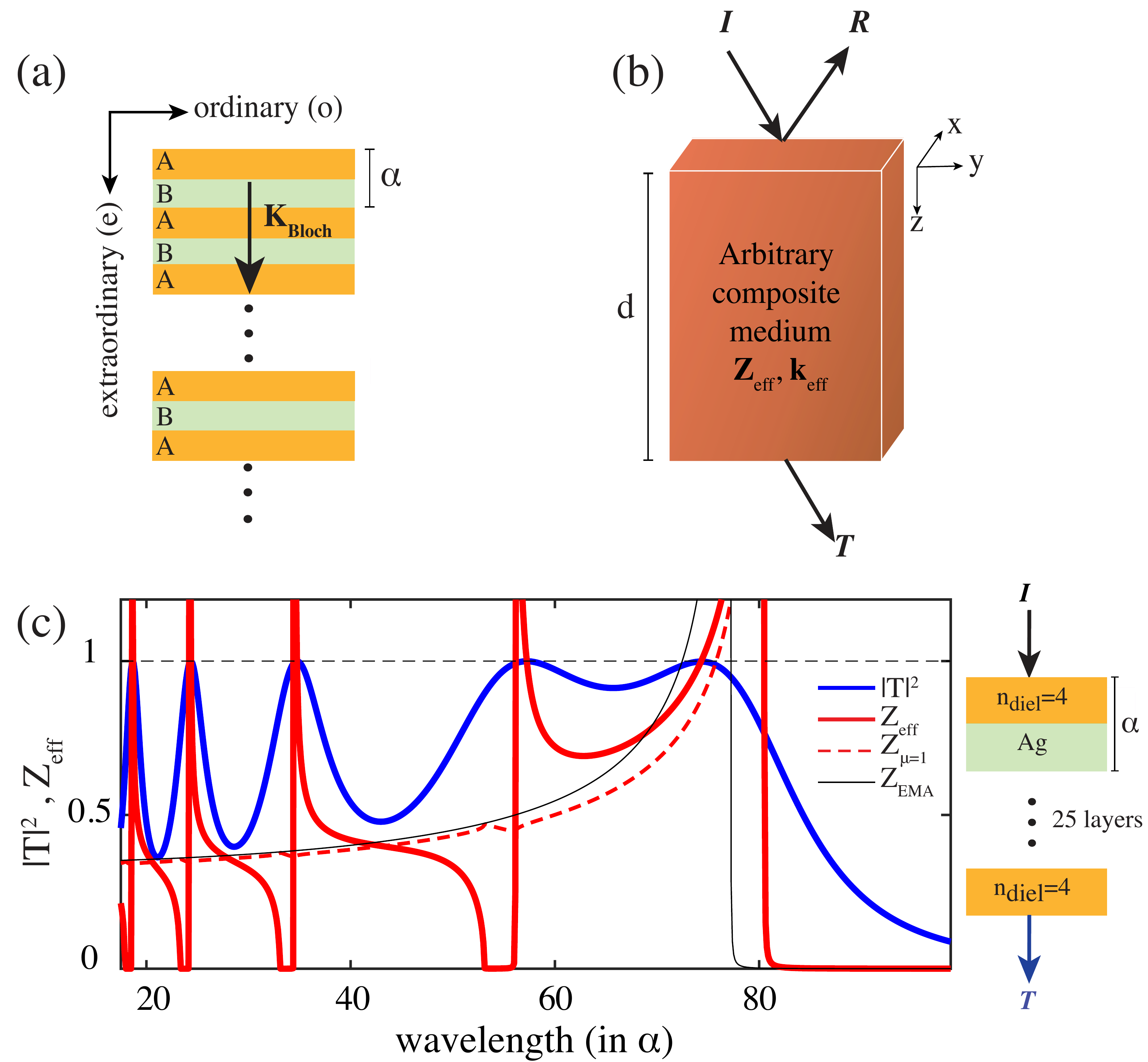}
  \caption{Comparison between (a) traditional EMA and Bloch approaches and (b) the general concept of S-parameter 
  retrievals. (c) Impedance-matching sanity check at normal incidence for a 25 layers dielectric/metal stack with 
  $n_{\mathrm{diel}}=4$. The transmittance $|T|^2$ calculation was performed with the 
  transfer-matrix formalism \cite{PochiYeh_Book} for the physical multilayer 
  system in the lossless limit. The dielectric and magnetic effective model ($\epsilon_{\mathrm{o}}$, $\mu_{\mathrm{o}}$) 
  accurately captures the structures resonances unlike the non-magnetic approach ($\epsilon_{\mathrm{o,\mu=1}}$) 
  and the Maxwell Garnett EMA.}
 \label{fig:Figure3}
\end{figure}

\par{\textcolor{black}{Contrary to the extensive use of EMAs, we utilize the S-parameter approach to describe 
dielectric/metal multilayer metamaterials of finite thickness. By letting the 
magnetic permeability $\mu_{\mathrm{eff}}$ be a free parameter, instead of \textit{a priori} 
setting $\mu_{\mathrm{eff}}=1$, we obtain magnetic resonances at wavelengths where 
magnetic dipole moments occur, as demonstrated in Figs.\ref{fig:Figure2}e, f. 
This confirms the physicality of the non-unity $\mu$; based on the arguments 
discussed in Sec.\ref{Concept}, magnetic resonances arise at wavelengths at which 
systems support circular or loop-like current distributions.}}

\par{\textcolor{black}{By accounting for the uniaxial anisotropy in planar heterostructures, we obtain both 
the ordinary and the extraordinary permeabilities $\mu_{\mathrm{o}}$ and $\mu_{\mathrm{e}}$, together with their 
dielectric permittivity counterparts, $\epsilon_{\mathrm{o}}$ and $\epsilon_{\mathrm{e}}$. As a sanity check, we first consider 
homogeneous metallic and dielectric slabs with known dielectric permittivity 
$\epsilon_{\mathrm{o}}=\epsilon_{\mathrm{e}}$ and $\mu_{\mathrm{o}}=\mu_{\mathrm{e}}=1$, which we recover upon 
application of our retrieval \cite{Papadakis_Retrieval}.}}

\par{\textcolor{black}{Another way to establish the validity of the the effective parameters is to perform an 
impedance-matching sanity check. Based on electromagnetic theory, the impedance of a structure 
at normal-incidence, $Z_{\mathrm{eff}}=\sqrt{\frac{\mu_{\mathrm{o}}}{\epsilon_{\mathrm{o}}}}$, must 
be unity at transmittance $|T|^2$ maxima. As seen in Fig. \ref{fig:Figure3}c, the retrieved 
parameters $\epsilon_{\mathrm{o}}$ and $\mu_{\mathrm{o}}$ accurately describe the scattering properties of planar dielectric/metal arrangements of 
finite thickness. On the contrary, not accounting for a magnetic permeability leads to inaccurate 
prediction of transmittance maxima. This is seen both by utilizing our S-retrieval-based approach while 
setting \textit{a priori} the magnetic permeability to unity ($Z_{\mathrm{\mu=1}}$ in Fig. \ref{fig:Figure3}c), 
and with the traditional EMA; both approaches fail to predict the 
structure's resonances.}}

\par{\textcolor{black}{By sweeping the angle of incidence from 0 to 90 degrees, in other words, by varying 
the in-plane wavenumbers $k_{\mathrm{//}}$, we obtain angle-independent \cite{Papadakis_Retrieval}, local material parameters 
for the systems we consider. This makes ellipsometry a suitable 
method to experimentally characterize our metamaterials in terms of local material tensorial 
parameters $\dvec{\mu}_{\mathrm{eff}}$ and $\dvec{\epsilon}_{\mathrm{eff}}$, as shown in the next 
section.
For larger $k_{\mathrm{//}}\gg\frac{\omega}{c}$, due 
to the plasmonic nature of the metallic layers, metal/dielectric arrangements exhibit some
degree of spatial dispersion \cite{Mote_Retrieval}. This effect is distinct from the magnetic resonances we 
investigate, which are the result of induced magnetic dipole moments (Sec.\ref{Concept}), consistent with
 the consensus in the field of artificial magnetism
\cite{Monticone_Magnetism}. Spatial dispersion is fully taken into account in what follows. This is 
done by extending our previous approach \cite{Papadakis_Retrieval} 
to consider as a free parameter not only the magnetic permeability, but also 
spatial dispersion in the form of 
wavenumber ($k_{\mathrm{//}}$) dependence (See discussion in  Sec.\ref{Implications} 
for Figs.\ref{fig:Figure5}b, c) \cite{Mote_Retrieval}. 
Furthermore, as seen by the experimentally confirmed effective parameters 
discussed in Fig. \ref{fig:Figure4}, all constituent permittivity and permeability 
components ($\epsilon_{\mathrm{o}}$, $\epsilon_{\mathrm{e}}$, $\mu_{\mathrm{o}}$, $\mu_{\mathrm{e}}$) are 
passive, causal, with positive imaginary parts and no antiresonance artifacts. Such artifacts are 
often associated with weak form of spatial dispersion (see \cite{Alu_Homogenization} 
and discussion \cite{Soukoulis_Antiresonances, Markel_NegativeImaginaryPart} 
among others).}}

\par{\textcolor{black}{Other approaches are also able to capture this artificial magnetic response
 by accounting for an effective permeability in 1D 
 metamaterials \footnote{\textcolor{black}{The general field-averaging scheme introduced by Smith and 
Pendry \cite{Pendry_Homogenization} is also able to capture the magnetic permeability
 we introduce, as discussed in Sec.\ref{Concept}. This scheme has been implemented in 
 the work by Watanabe et al., \cite{Watababe_TEBrewster}. Another approach can be found in 
 M. Iwanaga, Opt. Lett. \textbf{32}, 1314 (2007) and in references therein. Both approaches, however, 
 must be used with caution as they do not explicitly
account for spatial dispersion, contrary to \cite{Mote_Retrieval} and \cite{Papadakis_Retrieval} where
 spatial dispersion can be accounted for as a free parameter.}}.}}

\vspace{-1.0em}
\subsection{\label{Experimental Results}Experimental Results}
 \begin{figure}[ht!]
 \centering
\includegraphics[width=1.00\linewidth]{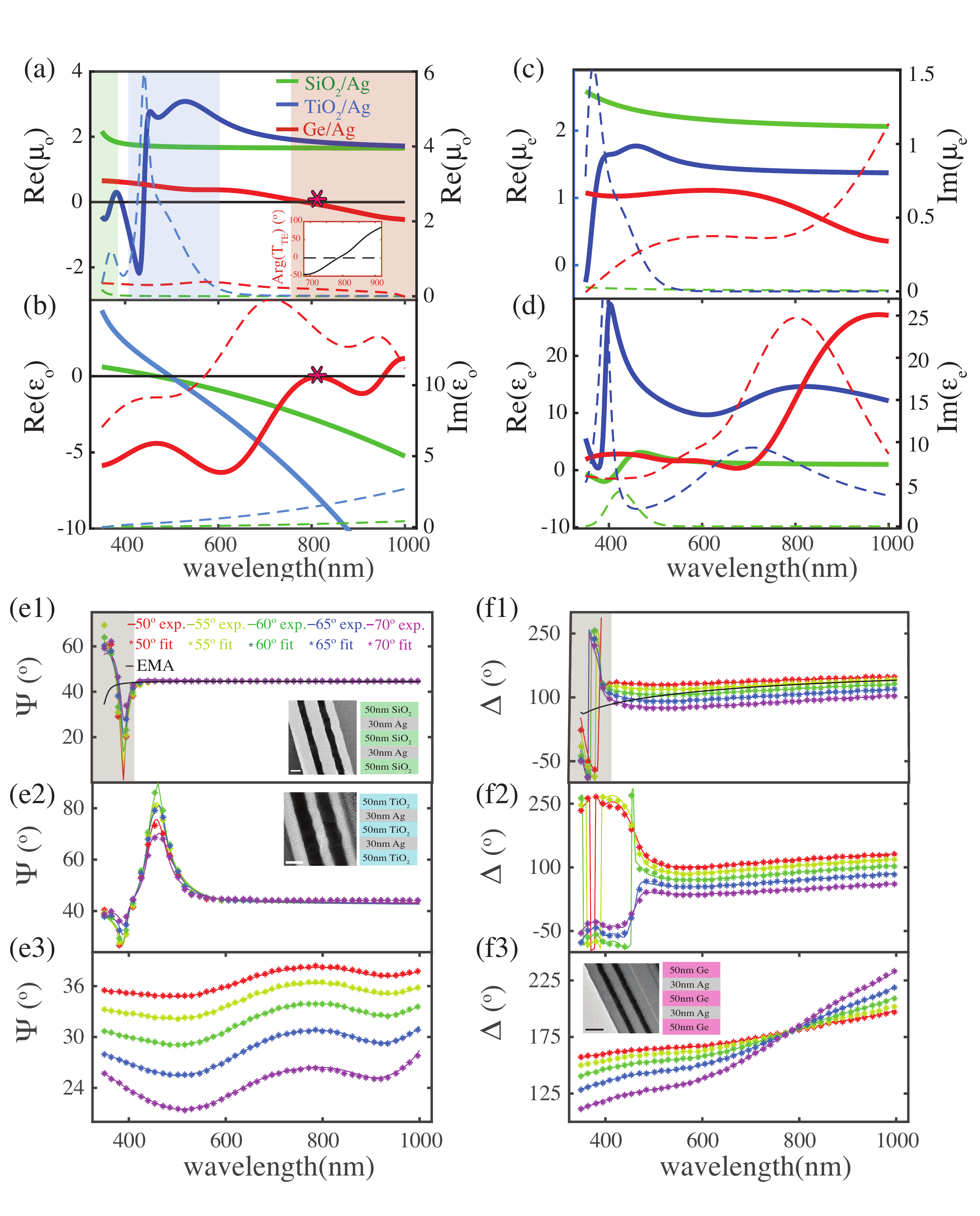}
  \caption{Experimentally determined (a) $\mu_{\mathrm{o}}$, (b) $\epsilon_{\mathrm{o}}$, (c) $\mu_{\mathrm{e}}$, (d) $\epsilon_{\mathrm{e}}$ for
   SiO\textsubscript{2}/Ag-green, TiO\textsubscript{2}/Ag-blue, Ge/Ag-red metamaterial. Solid lines-real
    parts, dashed lines-imaginary parts. Asterisks in (a) and (b): EMNZ wavelength for the Ge/Ag metamaterial, 
    inset in (a)-phase of transmission coefficient at the EMNZ wavelength. 
    (e1)-(e3) $\Psi$, (f1)-(f3) $\Delta$, solid lines-experiment, points-model 
    for: (e1), (f1): SiO\textsubscript{2}/Ag metamaterial, inset TEM scale bar: 50nm, 
  (e2), (f2): TiO\textsubscript{2}/Ag metamaterial, inset TEM scale bar: 50nm, 
  (e3), (f3): Ge/Ag metamaterial, inset TEM scale bar: 100nm }
 \label{fig:Figure4}
\end{figure}

\par{We fabricated multilayer structures by electron-beam evaporation and first measured the optical constants 
of the individual constituent layers with spectroscopic ellipsometry. We also determined their thicknesses 
with transmission electron microscopy (TEM). Thus, we were able to homogenize the layered metamaterials 
by assigning them effective parameters $\dvec{\epsilon}_{\mathrm{eff}}$ and $\dvec{\mu}_{\mathrm{eff}}$ using parameter 
retrieval methods \cite{Papadakis_Retrieval}, while taking into account fabrication imperfections. We 
then performed ellipsometric measurements of the full 
metamaterials and fit the experimental data with the \textcolor{black}{effective} parameters
 $\epsilon_{\mathrm{o}}$, $\epsilon_{\mathrm{e}}$, $\mu_{\mathrm{o}}$ and $\mu_{\mathrm{e}}$ 
in a uniaxial and Kramers-Kronig consistent model, \textcolor{black}{while fixing the total metamaterial thickness to the 
value determined through TEM imaging}. The fitting was over-determined as the number 
of incident angles exceeded the total number of fitting parameters.}

\par{The fabricated metamaterials were composed of SiO\textsubscript{2}/Ag, TiO\textsubscript{2}/Ag and Ge/Ag alternating 
layers (See TEM images and schematics in insets of Figs. \ref{fig:Figure4}e1, e2 and f3 respectively). The indices 
of the selected dielectric materials at optical frequencies 
are $n_{\mathrm{SiO_2}}\simeq1.5$, $n_{\mathrm{TiO_2}}\simeq2$ and $n_{\mathrm{Ge}}\simeq4-4.5$. As 
shown in Fig. \ref{fig:Figure4}a, increasing the dielectric index redshifts the magnetic resonance 
in the ordinary direction $\mu_{\mathrm{o}}$; the SiO\textsubscript{2}/Ag metamaterial supports a magnetic resonance 
in the long-wavelength UV regime ($\sim$300nm), whereas the TiO\textsubscript{2}/Ag and Ge/Ag metamaterials exhibit 
resonances in the blue (450nm) and red (800nm) part of the spectrum, respectively. 
The enhanced absorption in Ge at optical frequencies leads to considerable broadening of the 
Ge/Ag metamaterial magnetic resonance, yielding a broadband negative magnetic permeability 
for wavelengths above 800nm.}

\par{The presence of Ag induces a negative ordinary permittivity $\epsilon_{\mathrm{o}}$ (Fig. \ref{fig:Figure4}b) which, 
for the Ge/Ag metamaterial, becomes positive above 800nm due to the high-index of Ge. Notably, $\epsilon_{\mathrm{o}}$ 
crosses zero at 800nm, similar to $\mu_{\mathrm{o}}$, as emphasized with the asterisks in Figs. \ref{fig:Figure4}a, b. 
Thus, the Ge/Ag metamaterial exhibits an EMNZ response at optical frequencies. The EMNZ condition is confirmed by 
transfer-matrix analytical calculations of the physical multilayer structure. As shown in the inset of Fig. \ref{fig:Figure4}a),
 at the EMNZ wavelength, the phase of the transmission coefficient vanishes, showing that 
 electromagnetic fields propagate inside the metamaterial without phase advance \cite{Mahmoud_EMNZ}.}
 
\par{By comparing Fig. \ref{fig:Figure4}a to Fig. \ref{fig:Figure4}c one can infer that increasing the 
dielectric index leads to enhanced magnetic anisotropy. The 
parameter $\mu_{\mathrm{e}}$ only slightly deviates from $\mu_{\mathrm{o}}$ for the SiO\textsubscript{2}/Ag 
metamaterial, while the deviation is larger for the TiO\textsubscript{2}/Ag one. For the Ge/Ag 
metamaterial, $\mu_{\mathrm{e}}$ remains positive beyond 800nm, while $\mu_{\mathrm{o}}<0$, indicating \textit{magnetic} 
hyperbolic response for TE polarization. Furthermore, all three heterostructures exhibit hyperbolic 
response for TM polarization, with $\epsilon_{\mathrm{o}}<0$ and $\epsilon_{\mathrm{e}}>0$ (Figs.\ref{fig:Figure4}b, d). This 
makes the Ge/Ag metamaterial one with \textit{double} hyperbolic dispersion.}

 \par{The agreement between fitting and experimental data is very good, 
 as seen in Figs. \ref{fig:Figure4}e1-e3, f1-f3. \textcolor{black}{In Figs. \ref{fig:Figure4}e1, f1, we also provide 
 a Maxwell Garnett EMA-based fit for the  SiO\textsubscript{2}/Ag metamaterial. The
  EMA fails to reproduce the experimentally measured features, in both $\Psi$ and $\Delta$, that correspond to magnetic 
  permeability resonances (See grey-shaded region in Figs. \ref{fig:Figure4}e1, f1). Similar EMA-based 
  fits for the TiO\textsubscript{2}/Ag and Ge/Ag metamaterials 
  lead to large disagreement with the experimental data across the whole visible-near IR spectrum and are, thus, omitted. 
  This disagreement is expected, as the EMA approach is based on the assumption that the electric 
  field exhibits negligible or no variation within the lattice period \cite{Agranovich_EMA}, which does not 
  apply to high-index dielectric layers.}

\vspace{-1.5em}
\section{\label{Implications}BEYOND $\mu_{\mathrm{eff}}\neq 1$: FUNCTIONALIZING TE POLARIZATION IN PLANAR PHOTONICS}
 
\par{\textcolor{black}{In the previous sections we established, theoretically and experimentally, that
 dielectric/metal layered systems may be described with an effective magnetic permeability that deviates 
from unity. The purpose of introducing this parameter is to build a 
simple and intuitive approach for understanding and predicting new phenomena, for example, TE 
polarization response in planar systems. The non-unity and, in particular, the negative and anisotropic magnetic 
response that we demonstrated in Fig. \ref{fig:Figure4} motivates us to investigate TE characteristics of 
propagating modes (Figs.\ref{fig:Figure5}) and surface states (Fig. \ref{fig:Figure6}).
 We utilize an example system of dielectric/silver alternating layers, similar to the one we investigated
  experimentally. We let the refractive index of the dielectric material $n_{\mathrm{diel}}$ vary 
to emphasize that enhanced magnetic response at optical frequencies requires high-index dielectrics.}}
\par{\textcolor{black}{The calculations and full-wave simulations presented here are performed in the 
actual, physical, multilayer geometry (Figs.\ref{fig:Figure5}a, d, e and Fig.\ref{fig:Figure6}) and compared 
with the homogeneous effective slab picture 
($\dvec{\epsilon}_{\mathrm{eff}}$, $\dvec{\mu}_{\mathrm{eff}}$ - Figs.\ref{fig:Figure5}b, c). This helps assessing the 
validity of our model and emphasizing the physicality of the magnetic resonances.}}

\begin{figure}[]
 \centering
 \includegraphics[width=0.95\linewidth]{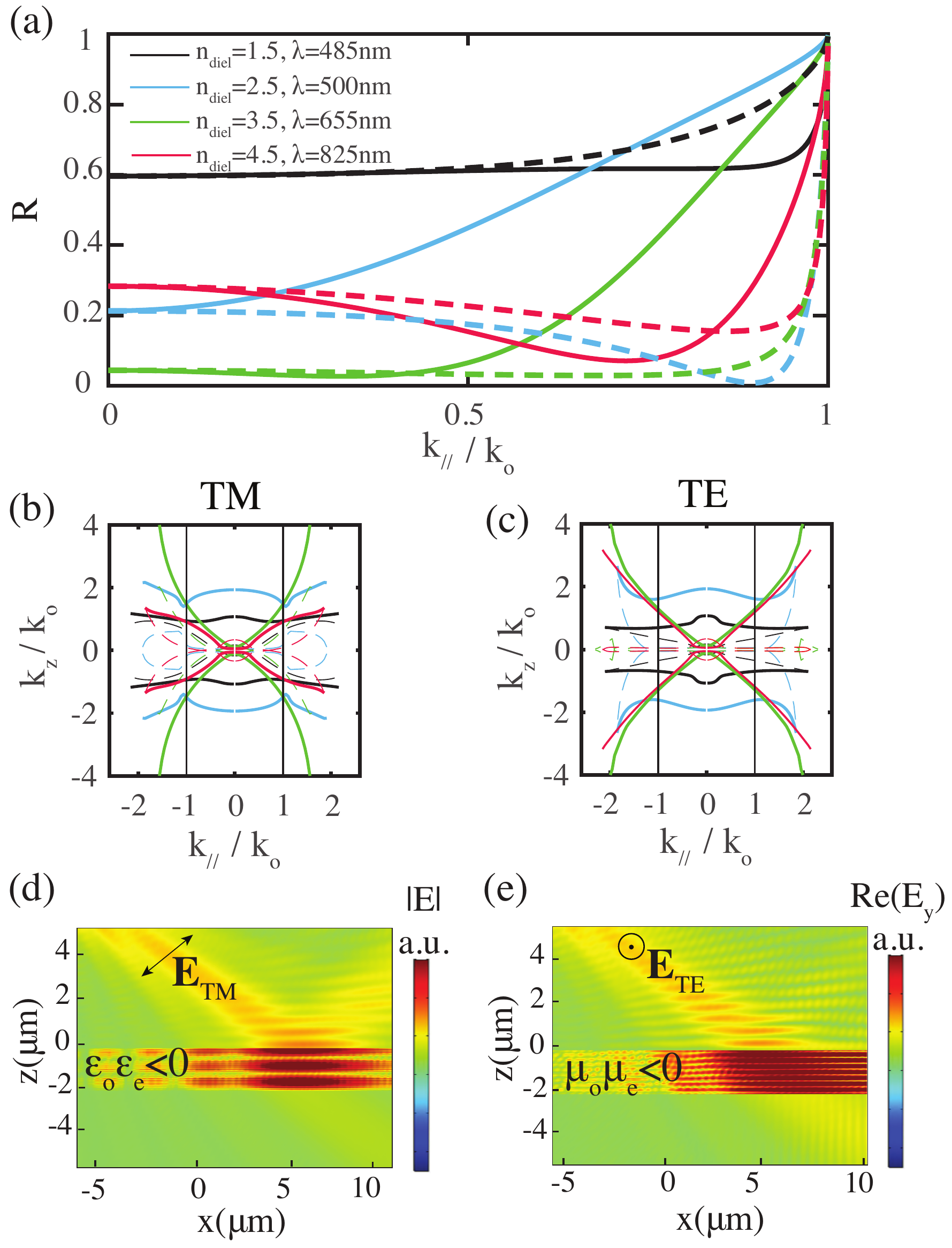}
  \caption{(a) Reflectance, solid lines: TE polarization, dashed lines: TM polarization, (b), (c) isofrequency 
  diagrams for TM and TE polarization, respectively, for a 5 layers dielectric $n_{\mathrm{diel}}$: 55nm/Ag: 25nm metamaterial.
  Solid lines: real parts, dashed lines: imaginary parts. 
  (d), (e): Simulation of a 55 layers dielectric $n_{\mathrm{diel}}=4$/Ag multilayer metamaterial. 
  The surrounding medium has index $n_{\mathrm{sur}}=1.55$, allowing coupling of high-k modes. We increased the 
  number of layers for clear visibility of field localization inside the structure. (d): TM polarization, (e): TE 
  polarization. Strong field localization is the consequence of (d) dielectric hyperbolic dispersion 
  for TM polarization ($\epsilon_{\mathrm{o}}\epsilon_{\mathrm{e}}<0$) and (e) magnetic hyperbolic 
  dispersion for TE polarization ($\mu_{\mathrm{o}}\mu_{\mathrm{e}}<0$). 
  }
 \label{fig:Figure5}
\end{figure}

\par{First, we perform transfer-matrix calculations for the example multilayer metamaterial and we show in Fig. \ref{fig:Figure5}a the 
angle dependence for TE and TM reflectance. The strong angle dependence of TM reflectance is 
well understood in the context of an equivalent homogeneous material with anisotropic effective 
dielectric response $\epsilon_{\mathrm{o}}\epsilon_{\mathrm{e}}<0$. Bulk TM modes experience dispersion}
\begin{equation} 
\frac{k_x^2+k_y^2}{\epsilon_{\mathrm{e}}(\omega,\vec{k})\mu_{\mathrm{o}}(\omega,\vec{k})} + \frac{k_z^2}{\epsilon_{\mathrm{o}}(\omega,\vec{k})\mu_{\mathrm{o}}(\omega,\vec{k})} = k_o^2
\end{equation}
where $k_o=\omega/c$. This dispersion is hyperbolic, as shown with isofrequency diagrams in Fig. \ref{fig:Figure5}b. Losses and spatial 
dispersion perturb the perfect hyperbolic shape \cite{Drachev_HMM}. In contrast to the TM modes, TE bulk 
modes interact with the magnetic anisotropy through the dispersion equation
\begin{equation} 
\frac{k_x^2+k_y^2}{\epsilon_{\mathrm{o}}(\omega,\vec{k})\mu_{\mathrm{e}}(\omega,\vec{k})} + \frac{k_z^2}{\epsilon_{\mathrm{o}}(\omega,\vec{k})\mu_{\mathrm{o}}(\omega,\vec{k})} = k_o^2
\end{equation}
 which is plotted in Fig. \ref{fig:Figure5}c. 
 For small wavenumbers ($k_{\mathrm{//}}/k_o<1$) and small dielectric indices $n_{\mathrm{diel}}$, the isofrequency diagrams 
 are circular, in other words, isotropic. This agrees well with our experimental results; as shown in 
 Figs. \ref{fig:Figure4}a, c, for the SiO\textsubscript{2}/Ag metamaterial,  ordinary and extraordinary 
 permeabilities do not drastically deviate from each 
other. Increasing the dielectric index opens the isofrequency contours, due to enhanced magnetic response in the 
ordinary direction ($\mu_{\mathrm{o}}$), which leads to magnetic anisotropy. We note that the displayed wavelengths
are selected at resonances of $\mu_{\mathrm{o}}$. Open TE polarization isofrequency contours for $n_{\mathrm{diel}}\geq2$ are 
consistent with both the experimental results (Fig. \ref{fig:Figure4}) for TiO\textsubscript{2} and Ge-based 
metamaterials and with the response of the physical multilayer structures, as shown in Fig. \ref{fig:Figure5}a; the 
TE reflectance indeed exhibits extreme angle dependence for increasing 
dielectric index. Strikingly, we observe a Brewster angle effect for TE polarization, which is 
unattainable in natural materials due to unity magnetic permeability at optical frequencies \cite{Watababe_TEBrewster}. 

\par{\textcolor{black}{An open isofrequency surface allows for coupling of large wavenumbers into a 
structure and enhancement in the density of optical states. This translates physically to 
strong interaction between incident light and a hyperbolic structure, and increased absorption, when it is possible to couple to large wavenumbers from the surrounding medium. So far,    
only TM polarization has been considered to experience this exotic hyperbolic response in planar 
dielectric/metal metamaterials, due 
to $\epsilon_{\mathrm{o}}\epsilon_{\mathrm{e}}<0$ \cite{Krishnamoorthy_TopTransitions, Drachev_HMM, Jacob_DOS}. Based on the 
open isofrequency surfaces for both TE and TM polarizations in Figs. \ref{fig:Figure5}b, 
c,  a high-index dielectric/metal multilayer metamaterial may
exhibit distinct frequency regimes of double, simultaneously TE and TM polarization, 
hyperbolic-like response. We perform
 finite element simulations of a  $n_{\mathrm{diel}}=4$/silver multilayer metamaterial for both linear polarizations 
 and set the index of the surrounding medium to  $n_{\mathrm{sur}}=1.55$ to allow coupling to 
larger wavenumbers. The well-known TM hyperbolic response is evident since the electric field is strongly 
localized within the multilayer in Fig. \ref{fig:Figure5}d. Switching the polarization
 to TE (Fig. \ref{fig:Figure5}e), we observe similar behavior, which, however, cannot be attributed to dielectric 
 anisotropic response as the electric field only experiences the in-plane dielectric permittivity ($\epsilon_{\mathrm{o}}$). 
 The TE enhanced absorption is associated with the $\mu_{\mathrm{o}}\mu_{\mathrm{e}}<0$ condition \cite{Dedkov_MagneticDOS}; 
 the number of TE modes supported by this metamaterial in this frequency regime is 
   drastically increased (See Supplementary Information).}}
   
  \par{Finally, we investigate surface wave propagation in our example system of a layered dielectric ($n_{\mathrm{diel}}$)/silver
   metamaterial. We do so by utilizing the transfer matrix mode condition $m_{\mathrm{11}}=0$ \cite{PochiYeh_Book}, which 
  we implement numerically using the reflection pole method \cite{Anemogiannis_RPM}. 
  In order to ensure interface-localized propagation with fields decaying in 
  air and in the metamaterial, we impose an additonal constraint for the states to be located in the optical 
  band gaps of both bounding media.}
  
  \par{For TM polarization, the identified surface states, with dispersion displayed in Fig. \ref{fig:Figure6}a, 
  bear similarity to typical SPPs on metallic interfaces \cite{Economou_SPP, Maier_Plasmonics} and to plasmonic waves in 
  metal/dielectric waveguides and systems \cite{Dionne_NegativeIndex}. Their plasmonic 
  nature is evident from their dispersion, which asymptotically approaches the surface plasma 
  frequency, similar to SPPs. We show in Fig. \ref{fig:Figure6}c their field distribution (dashed lines), and compare 
  to SPPs on an equivalent silver slab (black dotted lines). Such TM surface waves on metamaterial 
  interfaces are often associated with an effective negative dielectric 
  response \cite{Narimanov_Hypercrystal, Elser_AnisotropicPlasmon, Fang_TammMetalDiel}. 
  \textcolor{black}{This is consistent with our effective dielectric and magnetic model; as we explicitly
   showed experimentally in Fig. \ref{fig:Figure4}b, 
  the ordinary permittivity is negative $\epsilon_{\mathrm{o}}<0$, which results in TM plasmonic-like surface waves.}}

\begin{figure}[]
 \centering
 \includegraphics[width=1.03\linewidth]{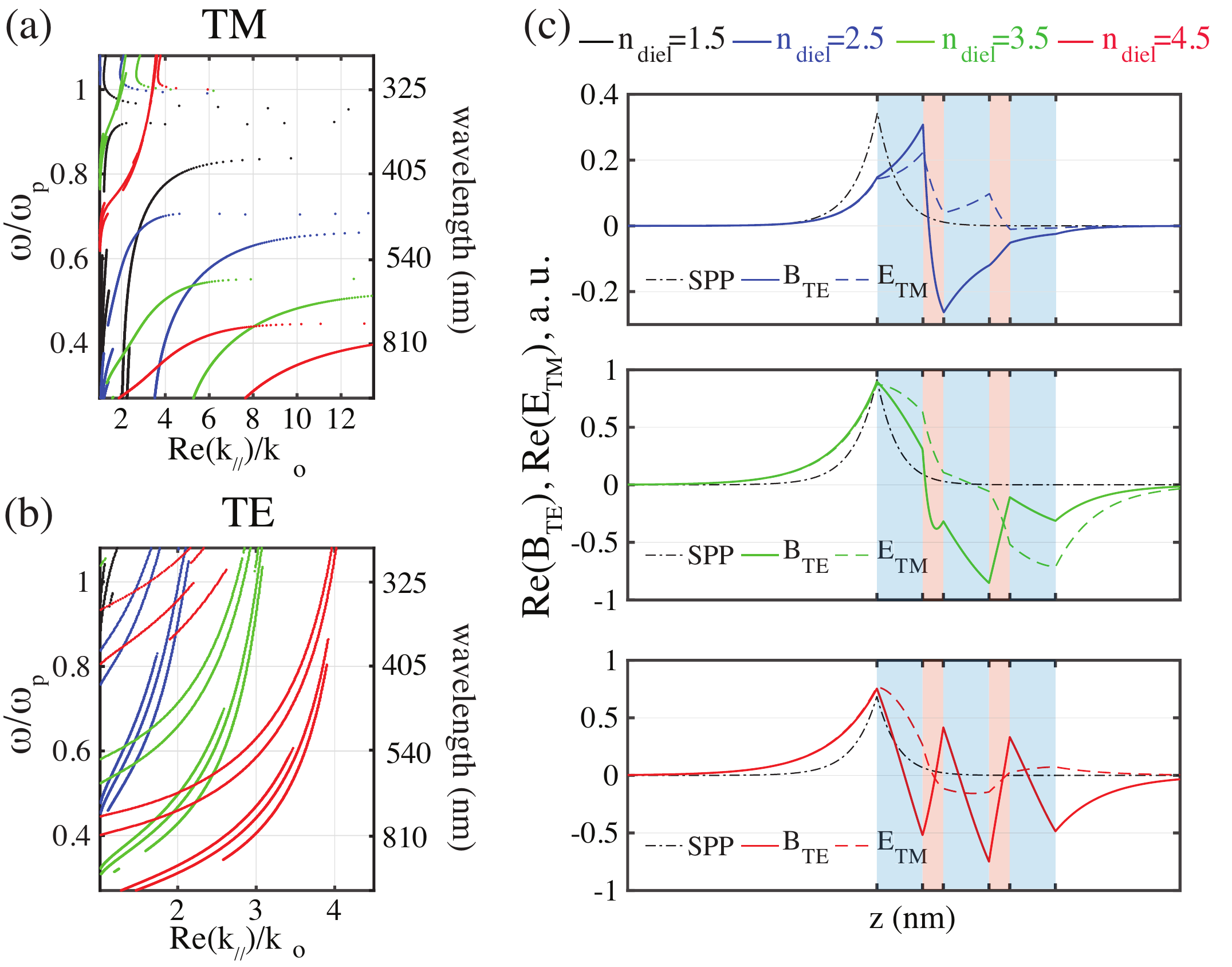}
  \caption{(a) TM and (b) TE surface wave dispersion for a 5 layers dielectric ($n_{\mathrm{diel}}$): 55nm/Ag: 25nm metamaterial. 
  (c) Field profiles (incidence from the left) and comparison to SPP mode on an equivalent Ag slab (black dotted line): 
  $n_{\mathrm{diel}}=2.5$, wavelength=620nm, $n_{\mathrm{diel}}=3.5$, wavelength=880nm, $n_{\mathrm{diel}}=4.5$, wavelength=1100nm.}
 \label{fig:Figure6}
\end{figure}

\par{Performing the same analysis for TE polarized waves, we find that TE surface-bound modes also exist (Fig. \ref{fig:Figure6}b). 
\textcolor{black}{Their dispersion is parabolic, resembling that of Tamm states 
in photonic crystals \cite{Fang_TammMetalDiel, Vinogradov_TETamm}. However, here, we show that they 
also exist in the subwavelength metamaterial limit and can coexist with typical TM plasmonic surface waves. 
TE polarized Tamm states have been previously theoretically associated only 
qualitatively with an arbitrary negative magnetic response \cite{Fang_TammMetalDiel}. 
By contrast, here, we explicitly connect their dispersion to values of magnetic permeabilities
 that were experimentally measured (Fig. \ref{fig:Figure4}). We identify their \textit{physical} 
origin, which is the strong displacement current supported in high-index dielectric layers 
with a loop-like distribution on resonance, as discussed in detail in Sec.\ref{Concept}. Specifically, from 
Fig. \ref{fig:Figure6}b, those TE surface waves emerge in the visible regime for dielectric layers with 
refractive index $n_{\mathrm{diel}}\geq2$ at frequencies where the metamaterial exhibits a negative effective magnetic 
response, which is also consistent with the empirical Eq.(\ref{eq:1}). For this 
reason, these states may be seen as ``magnetic plasmons''.}}

\par{The frequency regimes in which \textit{double} surface waves are supported demonstrate the 
 possibility of exciting TM polarized plasmonic modes simultaneously with their 
 TE counterparts in dielectric/metal pattern-free multilayers.}
  
\vspace{-1.5em}
\section{\label{sec:level1}CONCLUSIONS}

In conclusion, we have shown that non-unity effective magnetic permeability at optical frequencies 
can be obtained in one-dimensional layered systems, arising from displacement currents in dielectric 
layers. This makes it possible to tailor the magnetic response of planar HMMs, which have been previously
 only explored for their dielectric permittivity features. We experimentally demonstrated negative 
 in-plane magnetic permeability in planar structures, which can lead to double hyperbolic metamaterials. 
 By studying bulk and surface wave propagation, we have identified frequency regimes of a 
 rather polarization-insensitive response. We reported the existence of TE polarized 
 ``magnetic surface plasmons'', attributed to the negative effective magnetic permeability, 
 which are complementary to typical TM polarized surface plasmonic modes at the 
 interface of negative permittivity materials. The results reported here could open new directions 
 for tailoring wave propagation in artificial magnetic media in significantly simplified layered systems. 
 We anticipate that these findings will enable the generalization of the unique properties of plasmonics
  and hyperbolic metamaterials, previously only explored for TM polarized waves and negative 
  permittivity media, for unpolarized light at optical frequencies.}

This work was supported by U.S. Department of Energy (DOE) Office of Science grant 
DE-FG02-07ER46405 (A.D., D.F. and H.A.A.) and by the Multidisciplinary University 
Research Initiative Grant, Air Force Office of Scientific Research MURI, Grant No. 
FA9550-12-1-0488 (G.T.P). G.T. Papadakis acknowledges support by the National 
Science Foundation Graduate Research Fellowship, the American Association of 
University Women Dissertation Fellowship and NG Next at the Northrop Grumman Corporation. We 
acknowledge fruitful discussions with Dr. T. Tiwald, Dr. K. Thyagarajan, Dr. R. Pala, Dr. C. Santis and Dr. O. Ilic. We also thank 
C. Garland and B. Baker for assistance with sample preparation.

\bibliography{Manuscriptbib.bib}
\end{document}